\documentclass[aps,prl,superscriptaddress,twocolumn,floatfix]{revtex4-1}
\usepackage{graphicx,graphics}
\usepackage{dcolumn}
\usepackage{amsmath,amssymb,amsfonts}
\usepackage{latexsym,verbatim}
\usepackage{bm}
\usepackage{color}
\usepackage{ulem}
\usepackage[percent]{overpic}
\usepackage[breaklinks=true,colorlinks,citecolor=blue,linkcolor=blue,urlcolor=blue]{hyperref}

\begin{document}
\author{Alessandro Principi}
\affiliation{Radboud University, Institute for Molecules and Materials, NL-6525 AJ Nijmegen, The Netherlands}
\author{Mark B. Lundeberg}
\affiliation{ICFO-Institut de Ci\`encies Fot\`oniques, The Barcelona Institute of Science and Technology, 08860 Castelldefels (Barcelona), Spain}
\author{Niels C.H. Hesp}
\affiliation{ICFO-Institut de Ci\`encies Fot\`oniques, The Barcelona Institute of Science and Technology, 08860 Castelldefels (Barcelona), Spain}
\author{Klaas-Jan Tielrooij}
\affiliation{ICFO-Institut de Ci\`encies Fot\`oniques, The Barcelona Institute of Science and Technology, 08860 Castelldefels (Barcelona), Spain}
\author{Frank H.L. Koppens}
\affiliation{ICFO-Institut de Ci\`encies Fot\`oniques, The Barcelona Institute of Science and Technology, 08860 Castelldefels (Barcelona), Spain}
\affiliation{ICREA-Instituci\'{o} Catalana de Recer\c{a} i Estudis Avancats, Barcelona, Spain}
\author{Marco Polini}
\affiliation{Istituto Italiano di Tecnologia, Graphene Labs, Via Morego 30, I-16163 Genova, Italy}
\title{Super-Planckian electron cooling in a van der Waals stack}

\begin{abstract}
Radiative heat transfer (RHT) between macroscopic bodies at separations that are much smaller than the thermal wavelength is ruled by evanescent electromagnetic modes and can be orders of magnitude more efficient than its far-field counterpart, which is described by the Stefan-Boltzmann law. In this Letter we present a microscopic theory of RHT in van der Waals stacks comprising graphene and a natural hyperbolic material, i.e.~hexagonal boron nitride (hBN). We demonstrate that RHT between hot carriers in graphene and hyperbolic phonon-polaritons in hBN is extremely efficient at room temperature, leading to picosecond time scales for the carrier cooling dynamics.
\end{abstract}

\maketitle

{\it Introduction.---}The cooling stages of the temperature dynamics of hot carriers in a crystal typically proceed via energy transfer to phonons~\cite{hotcarriers}. In the case of pristine graphene, ultra-long cooling times, on the order of nanoseconds, have been theoretically predicted~\cite{bistritzer_prl_2009,tse_prb_2009}. Such slow cooling dynamics is due to energy transfer to graphene acoustic phonons via collisions that conserve momentum. If realized experimentally, this intrinsic relaxation dynamics would imply notable figures of merit for graphene-based photodetectors~\cite{koppens_naturenano_2014}. Unfortunately, the cooling dynamics in ``first-generation'' graphene samples~\cite{geim_naturemater_2007}, i.e.~samples deposited on ${\rm SiO}_2$, is believed to be dominated by far more efficient disorder-assisted {\it momentum-non-conserving} collisions between electrons and graphene acoustic phonons~\cite{betz_naturephys_2012,graham_naturephys_2013,song_prl_2012,song_jpcm_2015}. According to theory~\cite{song_prl_2012,song_jpcm_2015}, such ``supercollisions" are due to short-range (rather than long-range) disorder. 

It is therefore not clear yet how to reach the intrinsic regime~\cite{bistritzer_prl_2009,tse_prb_2009}. In this respect, a natural question arises: What is the fate of the temperature dynamics of hot carriers in ``second-generation'' samples~\cite{geim_nature_2013}, where graphene is encapsulated between hexagonal boron nitride (hBN) crystals~\cite{mayorov_nanolett_2011,mayorov_nanolett_2012,wang_science_2013,taychatanapat_naturephys_2013,woessner_naturemater_2015,bandurin_science_2016}? On the one hand, these samples have shown nearly ideal transport characteristics~\cite{mayorov_nanolett_2011,mayorov_nanolett_2012,wang_science_2013,taychatanapat_naturephys_2013,woessner_naturemater_2015,bandurin_science_2016}, whereby {\it momentum-conserving} electron-acoustic phonon scattering~\cite{hwang_prb_2008,principi_prb_2014} fully determines dc transport times at room temperature, at least for sufficiently large carrier densities. On the other hand, hBN crystal slabs are known to support low-loss standing Fabry-P\'erot phonon-polaritons~\cite{dai_science_2014,caldwell_naturecommun_2014,tomadin_prl_2015}. These modes occur because hBN is a uniaxial crystal with intrinsic {\it hyperbolic} character~\cite{hyperbolicmaterials}, i.e.~with in- ($\epsilon_x$) and out-of-plane ($\epsilon_z$) components of the dielectric tensor $\hat{\bm \epsilon}$ having opposite signs in the so-called ``reststrahlen'' frequency bands. 

Could radiative heat transfer (RHT) to hyperbolic phonon-polaritons in hBN significantly affect the late stages of the cooling dynamics of hot carriers in graphene? In this Letter we answer this question affirmatively. 

RHT between macroscopic bodies has been studied since they early days of 1900, when Planck explained the black-body radiation spectrum. In the regime $d \gg d_{T}$, where $d$ is the separation between two bodies and $d_{T} = \hbar c/(k_{\rm B} T)$ is the thermal wavelength, RHT is due to traveling electromagnetic (EM) waves and is controlled by the Stefan-Boltzmann law. On the contrary, in the limit $d \ll d_{T}$ RHT is dominated by evanescent modes of the EM field and power transfer can greatly exceed the black-body limit (``super-Planckian'' thermal emission). Anomalous RHT between closely spaced bodies was first studied experimentally by Hargreaves~\cite{hargreaves_pla_1969}. This pioneering work motivated the introduction of a general theoretical formalism~\cite{polder_prb_1971}, which was based on the theory of fluctuating electrodynamics~\cite{rytov_book_1953,lifshitz_zetf_1956,dzyaloshinskii_ap_1961}. Near-field thermal coupling has been intensively studied in the past both experimentally~\cite{wilde_nature_2006,shen_nanolett_2009,rousseau_naturephoton_2011,ottens_prl_2011,jones_nanolett_2012,kim_nature_2015} and theoretically~\cite{pendry_jpcm_1999,volokitin_prb_2001,benabdallah_prb_2010,guo_apl_2012,messina_prb_2013,benabdallah_prl_2014,miller_prl_2015,liu_acsphoton_2014}. 

In this work we present a theoretical study of RHT between a two-dimensional (2D) gas of hot massless Dirac fermions (MDFs) in graphene~\cite{kotov_rmp_2012} and nearby slabs of hBN crystals. We follow an approach that differs from fluctuating electrodynamics~\cite{polder_prb_1971,rytov_book_1953,lifshitz_zetf_1956,dzyaloshinskii_ap_1961} and relies on a combination of Fermi's golden rule with an analytic electrostatic calculation of how phonon-polaritons modify the instantaneous free-space photon propagator. We have, however, checked (not shown in this work) that Eqs.~(\ref{eq:Q_rate_fin})-(\ref{eq:Z_factor}) can also be obtained from fluctuating electrodynamics. We demonstrate that the hyperbolic nature of hBN crystals sets an extremely efficient intrinsic pathway for the dissipation of heat stored by graphene carriers. Cooling into non-hyperbolic polar substrates has been studied e.g.~in Ref.~\onlinecite{low_prb_2012}. Below, we set $\hbar=k_{\rm B} =1$, unless explicitly stated otherwise.

{\it Electron-photon coupling and the hBN-dressed photon propagator.---}RHT occurs because of the coupling between MDFs in graphene and the surrounding three-dimensional (3D) EM field. This coupling is described 
by the usual light-matter interaction Hamiltonian
\begin{eqnarray}\label{eq:eph_Hamiltonian}
{\cal H}_{\rm eph} = \frac{e}{c V} \sum_{{\bm q}, q_z, \nu} {\bm j}_{-{\bm q}}\cdot {\bm A}_{{\bm q}, q_z, \nu}(t)~,
\end{eqnarray}
where $-e$ is the electron charge, $c$ is the speed of light in vacuum, $V$ is the 3D quantization volume, $\nu = {\rm TM}, {\rm TE}$ is the polarization of the EM field, and ${\bm j}_{\bm q}= \sum_{\bm k, \lambda, \lambda^\prime} c^\dagger_{{\bm k} - {\bm q}, \lambda} {\bm J}_{{\bm k} - {\bm q}, \lambda; {\bm k}, \lambda'} c_{{\bm k}, \lambda^\prime}$ is the MDF particle current operator~\cite{kotov_rmp_2012}. Here, $c_{{\bm k},\lambda}^\dagger$ ($c_{{\bm k},\lambda}$) creates (destroys) an electron with momentum ${\bm k}$ in band $\lambda = \pm$, and ${\bm J}_{{\bm k}, \lambda; {\bm k}^\prime, \lambda'}$ are the matrix elements of the current operator in the band representation. The intra-band matrix element is ${\bm J}_{{\bm k},\lambda;{\bm k}',\lambda} = \lambda v_{\rm F} \big[ \cos(\theta_{{\bm k},{\bm k}'}/2), \sin(\theta_{{\bm k},{\bm k}'}/2) \big]$, while the inter-band one is ${\bm J}_{{\bm k},\lambda;{\bm k}',{\bar \lambda}} = i {\hat {\bm z}} \times {\bm J}_{{\bm k},\lambda;{\bm k}',\lambda}$. Here $\theta_{{\bm k},{\bm k}'} \equiv \theta_{{\bm k}} + \theta_{{\bm k}'}$ is the sum of the polar angles of the vectors ${\bm k}$ and ${\bm k}^{\prime}$ and ${\bar \lambda} = -\lambda$.  Note that the current operator 
${\bm j}_{\bm q}$ (photon field ${\bm A}_{{\bm q}, q_z, \nu}$) in Eq.~(\ref{eq:eph_Hamiltonian}) is represented by a 2D (3D) vector. 

We now introduce the matrix element of the light-matter interaction,
\begin{equation}\label{eq:matrix_element}
u_{{\bm k}, {\bm q}, q_{z}}^{\lambda,\lambda',\nu} = \sqrt{\frac{2\pi e^2}{\omega_{{\bm q}, q_{z},\nu}}} {\bm e}_\nu\cdot{\bm J}_{{\bm k},\lambda;{\bm k}+{\bm q},\lambda'}
\end{equation}
where $\omega_{{\bm q}, q_{z},\nu}$ are the frequencies of the photonic modes, and the second-quantized expression for the Fourier components of the photon field,
\begin{equation}\label{eq:vector_potential_def}
{\bm A}_{{\bm q}, q_{z}, \nu}(t) = \sqrt{\frac{2\pi c^2}{\omega_{{\bm q}, q_{z}, \nu}}} {\bm e}_\nu 
[a^\dagger_{{\bm q}, q_{z}, \nu}(t) + a_{-{\bm q}, - q_{z}, \nu}(t)]~,
\end{equation}
where $a_{{\bm q},\nu}^\dagger(t)$ ($a_{{\bm q},\nu}(t)$) creates (destroys) a photon with momentum ${\bm q}$ and polarization $\nu$ at time $t$. The corresponding 3D photon propagator
\begin{equation}\label{eq:propagator_real_time}
G^{\rm 3D}_{\nu,\alpha\beta}({\bm q},q_z,t) \equiv -i \frac{e^2}{c^2}\langle T A_{{\bm q},q_{z}, \nu, \alpha}(t)A^\dagger_{{\bm q}, q_{z}, \nu,\beta} \rangle
\end{equation}
and its Fourier transform
\begin{equation}\label{eq:propagator_FT}
G^{\rm 3D}_{\nu,\alpha\beta}({\bm q},q_z,\omega) \equiv \int_{-\infty}^{+\infty}dt~e^{i\omega t}G^{\rm 3D}_{\nu,\alpha\beta}({\bm q},q_z,t)
\end{equation}
contain all the necessary information. Here, ``$T$'' denotes the time-ordering operator, $\alpha,\beta=x,y,z$ are Cartesian indices, and $\langle\dots\rangle$ denotes an average over the thermal ensemble. For photons in {\it free space}, the propagator (\ref{eq:propagator_FT}) reduces to
\begin{equation} \label{eq:bare_photon_prop}
G^{0, {\rm 3D}}_{\nu,\alpha\beta}({\bm q}, q_z, \omega) =
-\frac{4\pi e^2}{(\omega+i\eta)^2 - \omega_{{\bm q}, q_{z}, \nu}^2}  e_{\nu,\alpha} e_{\nu,\beta}~,
\end{equation}
where $e_{\nu,\alpha}$ is the $\alpha$-th component of the polarization vector corresponding to the photonic mode $\nu$. 

Because of the phonons in the nearby hBN crystals, however, the bare propagator (\ref{eq:bare_photon_prop}) is substantially altered. This ``dressing'' can be easily captured analytically in the non-retarded limit, where we can use the following relation between the vector potential and the instantaneous electrostatic potential: $\partial_t {\bm A}_{{\bm q}, q_{z}, \nu} = i c {\bm q} \phi({\bm q}, q_{z}, t)$, for $\nu={\rm TM}$. This identity implies that, in the non-retarded limit, the 3D dressed propagator $G^{\rm 3D}_{\nu,\alpha\beta}({\bm q},q_z,t)$ can be calculated from the knowledge of the 3D instantaneous Coulomb propagator. The latter, in turn, can be calculated by utilizing a straightforward electrostatic approach described in Ref.~\onlinecite{tomadin_prl_2015}. Finally, the required 2D propagator can be obtained after integrating over $q_{z}$. The end result is
\begin{equation}\label{eq:final_propagator}
{\rm Im}[G^{\rm 2D}_{\nu,\alpha,\beta}({\bm q},\omega)] =
\left\{
\begin{array}{l}
{\displaystyle \frac{q_\alpha q_\beta}{\omega^2} \Im m\left[V_{{\bm q},\omega}\right],~{\rm if}~\nu = {\rm TM}}
\vspace{0.2 cm}\\
0,~{\rm if}~\nu = {\rm TE}
\end{array}
\right.~,
\end{equation}
where $V_{{\bm q},\omega}$ is the instantaneous Coulomb propagator dressed by the presence of the surrounding hBN dielectric slabs. Its frequency dependence stems from the frequency dependence of the in-plane and out-of-plane permittivities $\epsilon_{x}(\omega)$ and $\epsilon_{z}(\omega)$ of hBN. Explicit expressions for $V_{{\bm q},\omega}$ and the permittivities $\epsilon_{x}(\omega)$ and $\epsilon_{z}(\omega)$ are reported in Eqs.~(\ref{eq:effective_interaction_uniaxial})-(\ref{eq:BN_dielectric}) below. As discussed in Ref.~\onlinecite{tomadin_prl_2015}, poles of the dressed Coulomb interaction $V_{{\bm q},\omega}$ yield the dispersion relation of standing phonon-polariton modes in the hBN slabs surrounding graphene. We now proceed to calculate RHT between hot carriers in graphene and the dressed EM field around graphene.

{\it Boltzmann-transport theory of RHT.---}The calculation of the 2D dressed  propagator (\ref{eq:final_propagator}) allows us to calculate how phonon-polaritons dress the squared matrix element of the light-matter interaction. We start by squaring the bare matrix element in Eq.~(\ref{eq:matrix_element}):
\begin{eqnarray}\label{eq:squared_matrix_element_bare}
\left| u_{{\bm k}, {\bm q}, q_{z}}^{\lambda,\lambda',\nu} \right|^2 &=&\frac{2\pi e^2}{\omega_{{\bm q}, q_{z},\nu}}\sum_{\alpha,\beta =x,y} J^{(\alpha)}_{{\bm k},\lambda;{\bm k}+{\bm q},\lambda'}  J^{(\beta)}_{{\bm k}+{\bm q},\lambda';{\bm k},\lambda} e_{\nu,\alpha} e_{\nu,\beta}\nonumber\\
&=&- \sum_{\alpha,\beta =x,y} J^{(\alpha)}_{{\bm k},\lambda;{\bm k}+{\bm q},\lambda'}  J^{(\beta)}_{{\bm k}+{\bm q},\lambda';{\bm k},\lambda}\nonumber\\
&\times&\int_0^\infty \frac{d\omega}{\pi} {\rm Im}[G^{0, {\rm 3D}}_{\nu,\alpha\beta}({\bm q}, q_{z},\omega)]~.
\end{eqnarray}
Coupling of MDFs to phonon-polaritons is achieved by replacing the bare propagator $G^{0, {\rm 3D}}_{\nu,\alpha\beta}({\bm q}, q_{z},\omega)$ in Eq.~(\ref{eq:squared_matrix_element_bare}) with the dressed propagator $G^{\rm 3D}_{\nu,\alpha\beta}({\bm q}, q_{z},\omega)$. Following this procedure and integrating over $q_{z}$, we find an effective 2D dressed squared matrix element:
\begin{eqnarray}\label{eq:final_result_squared_matix_element}
\left| U_{{\bm k},{\bm q}}^{\lambda,\lambda',\nu} \right|^2 &\equiv&
\int_{-\infty}^{+\infty}\frac{dq_{z}}{2\pi} \left| U_{{\bm k}, {\bm q}, q_{z}}^{\lambda,\lambda',\nu} \right|^2 \nonumber\\
&=& -\sum_{\alpha,\beta =x,y} J^{(\alpha)}_{{\bm k},\lambda;{\bm k}+{\bm q},\lambda'}  J^{(\beta)}_{{\bm k}+{\bm q},\lambda';{\bm k},\lambda}\nonumber\\
&\times&\int_0^\infty \frac{d\omega}{\pi} {\rm Im}[G^{\rm 2D}_{\nu,\alpha\beta}({\bm q},\omega)]
\end{eqnarray}
We now feed Eq.~(\ref{eq:final_result_squared_matix_element}) to the collision integral 
${\cal I}_{{\bm k}, \lambda}$ in  a semiclassical Boltzmann equation $\partial_t f_{{\bm k},\lambda} = -{\cal I}_{{\bm k},\lambda}$ for the electron distribution function $f_{{\bm k},\lambda}$~\cite{bistritzer_prl_2009,tse_prb_2009,song_prl_2012,song_jpcm_2015}. Here, 
\begin{eqnarray}
{\cal I}_{{\bm k},\lambda} &=& \sum_{{\bm k}',\lambda'} \big[ f_{{\bm k},\lambda} (1 - f_{{\bm k}',\lambda'}) W_{{\bm k},\lambda\to {\bm k}',\lambda'} -  f_{{\bm k}',\lambda'} (1 - f_{{\bm k},\lambda}) \nonumber\\
&\times&W_{{\bm k}',\lambda'\to {\bm k},\lambda} \big]
~,
\end{eqnarray}
and the transition probability
\begin{eqnarray}\label{eq:W}
W_{{\bm k},\lambda\to {\bm k}',\lambda'} &=& 2\pi \sum_{{\bm q},\nu} \left| U_{{\bm k},{\bm q}}^{\lambda,\lambda',\nu} \right|^2 \big[ 
(n_{{\bm q},\nu} + 1) \delta(\Delta \varepsilon - \omega_{{\bm q},\nu}) \nonumber\\
&\times& \delta({\bm k} - {\bm k}' - {\bm q})
+ n_{{\bm q},\nu} \delta(\Delta \varepsilon + \omega_{{\bm q},\nu}) \nonumber\\
&\times&\delta({\bm k} - {\bm k}' + {\bm q})
\big]~,
\end{eqnarray}
where $n_{{\bm q,\nu}}$ is the phonon-polariton distribution function and $\Delta \varepsilon = \varepsilon_{{\bm k},\lambda} - \varepsilon_{{\bm k}',\lambda'}$ the electronic transition energy.  

We are now in the position to calculate the energy transfer rate, which we define to be positive if $T_{\rm e}>T_{\rm L}$. Multiplying both members of the Boltzmann equation by $\varepsilon_{{\bm k},\lambda} - \mu$ and summing over ${\bm k}, \lambda$ we find an equation of motion for the energy density ${\cal E}$:
\begin{equation}\label{eq:EOM}
\partial_{t}{\cal E} = - \tilde{\cal Q}~,
\end{equation}
where the energy transfer rate is given by
\begin{eqnarray}\label{eq:Q_rate_def}
\tilde{\cal Q} &=& -\sum_{\nu} \int\frac{d^2{\bm q}}{(2\pi)^2}\int_{-\infty}^\infty \frac{d\omega}{\pi} \omega 
 \big[ n_{\rm B}(\omega/T_{\rm L}) - n_{\rm B}(\omega/T_{\rm e}) \big] \nonumber\\
 &\times&{\rm Im}[\chi^{(0)}_{j_\alpha,j_\beta}({\bm q},\omega)]~{\rm Im}[G^{\rm 2D}_{\nu,\alpha\beta}({\bm q},\omega)]~.
\end{eqnarray}
Here, $\chi^{(0)}_{j_\alpha,j_\beta}({\bm q},\omega)$ is the current-current response tensor of a 2D system of {\it non-interacting} MDFs~\cite{principi_prb_2009}. To obtain Eq.~(\ref{eq:Q_rate_def}) we assumed that both electrons and phonon-polaritons are at equilibrium at the two temperatures 
$T_{\rm e}$ (electron temperature) and $T_{\rm L}$ (lattice temperature), respectively. Therefore, $f_{{\bm k},\lambda}$ and $n_{{\bm q,\nu}}$ are equilibrium distribution functions: $f_{{\bm k},\lambda} \equiv n_{\rm F}((\varepsilon_{{\bm k},\lambda}-\mu)/T_{\rm e})$ and $n_{{\bm q,\nu}} = n_{\rm B}(\omega_{{\bm q},\nu}/T_{\rm L})$, where $n_{{\rm F}, {\rm B}}(x) = (e^x\pm 1)^{-1}$ is the Fermi-Dirac (Bose-Einstein) distribution. The chemical potential $\mu = \mu(T_{\rm e})$ is obtained by requiring the particle density $n$ to be time-independent.

\begin{figure}[t]
\begin{center}
\begin{overpic}[width=0.9\columnwidth]{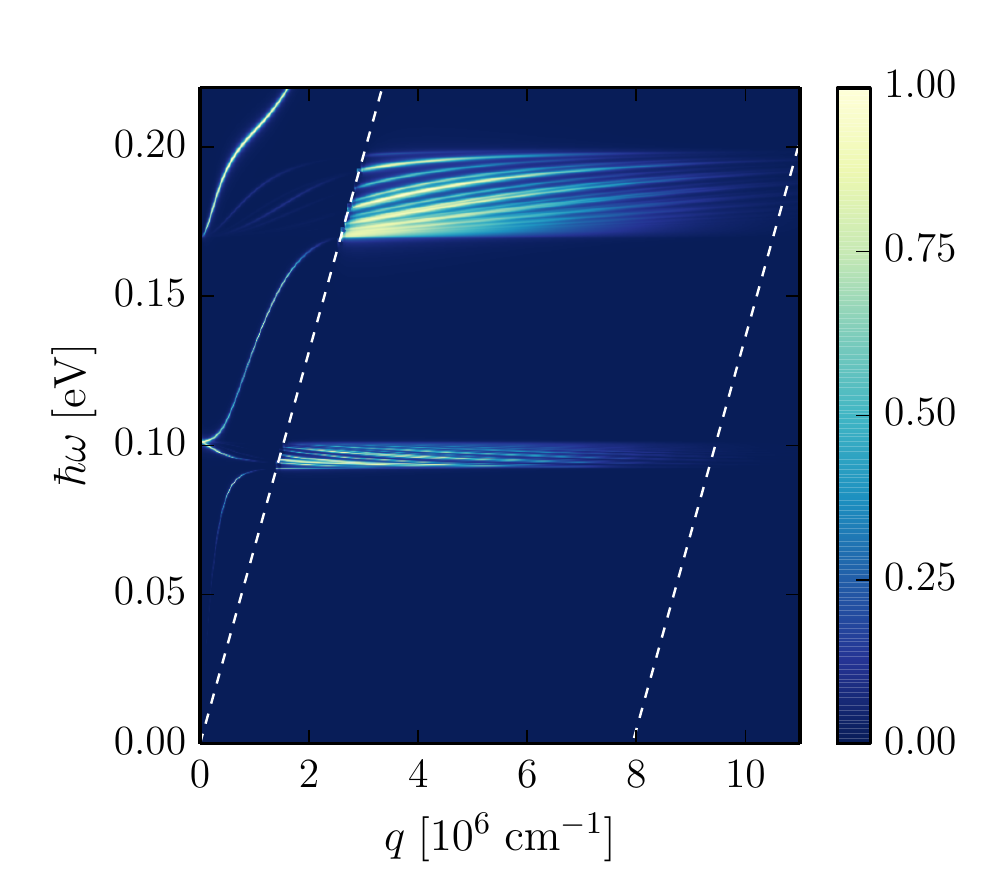}\put(2,80){(a)}
\end{overpic}\\
\begin{overpic}[width=0.9\columnwidth]{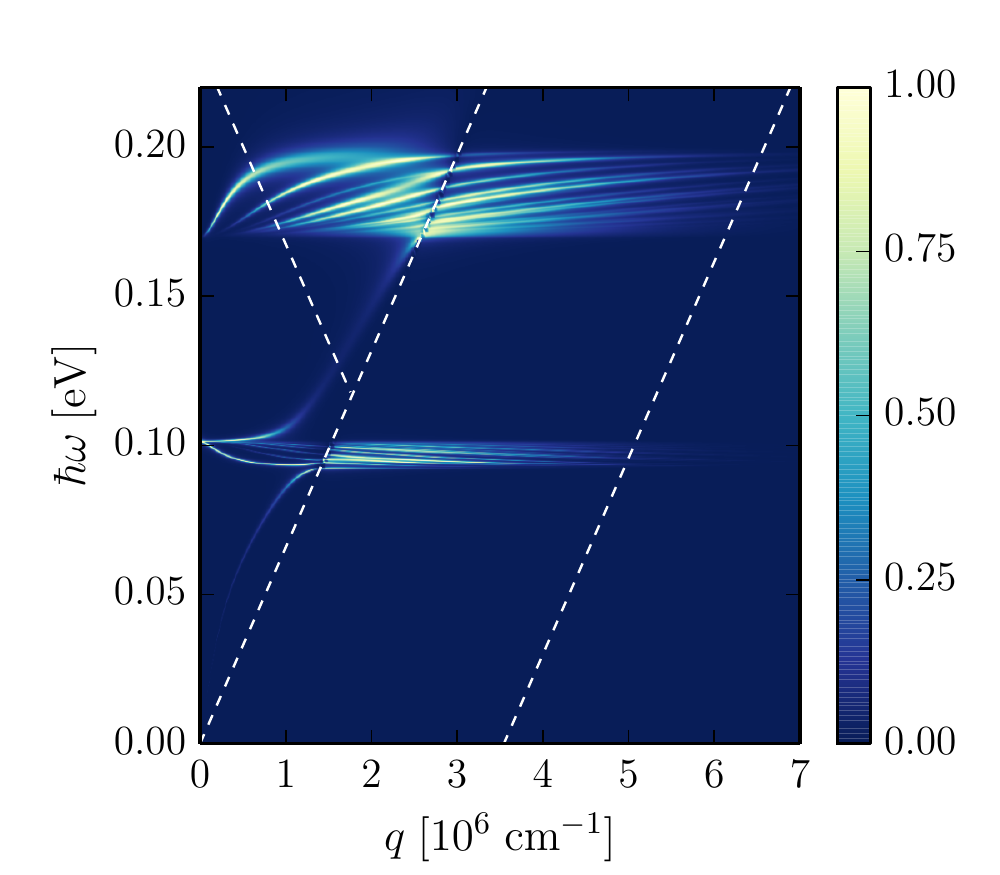}\put(2,80){(b)}
\end{overpic}\\
\end{center}
\caption{(Color online) Color plots of the dimensionless function ${\cal Z}(q,\omega)$ in Eq.~(\ref{eq:Z_factor}) for an electron temperature $T_{\rm e}=300~{\rm K}$ and a top (bottom) hBN thickness 
$d^\prime = 9~{\rm nm}$ ($d = 27~{\rm nm}$). All other heterostructure and hBN parameters (phonon frequencies and lifetimes) are reported in the Appendix below. Panel (a) $n=5.0 \times 10^{12}~{\rm cm}^{-2}$. Panel (b) $n=1.0 \times 10^{12}~{\rm cm}^{-2}$. Dashed lines indicate the edges of the graphene intra-band and inter-band particle-hole continua~\cite{kotov_rmp_2012}. In panel (a), the bottom edge of the inter-band continuum is not present since it occurs well above the hBN reststrahlen bands for $n=5\times 10^{12}~{\rm cm}^{-2}$. In both panels, we clearly see that ${\cal Z}(q,\omega)$ is maximum at the location of the poles of the dressed Coulomb interaction $V_{{\bm q}, \omega}$ (standing hBN phonon-polariton modes~\cite{dai_science_2014,caldwell_naturecommun_2014}) and at the zeroes of the dynamical dielectric function $\varepsilon(q,\omega)$ (plasmon-phonon polariton branches~\cite{tomadin_prl_2015,woessner_naturemater_2015}).\label{fig:one}}
\end{figure}

\begin{figure}[t]
\begin{center}
\begin{tabular}{c}
\begin{overpic}[width=1.0\columnwidth]{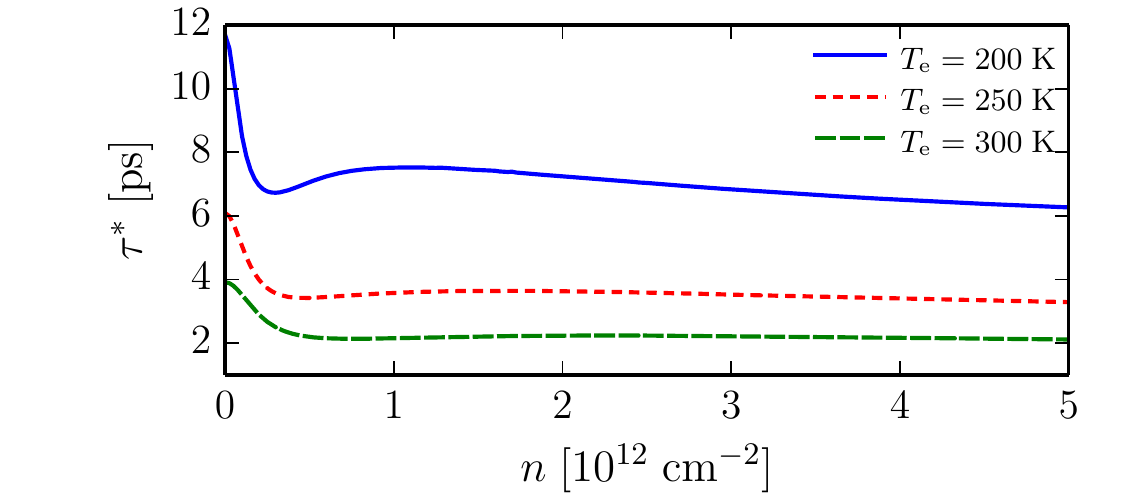}\put(2,42){(a)}
\end{overpic}\\
\begin{overpic}[width=1.0\columnwidth]{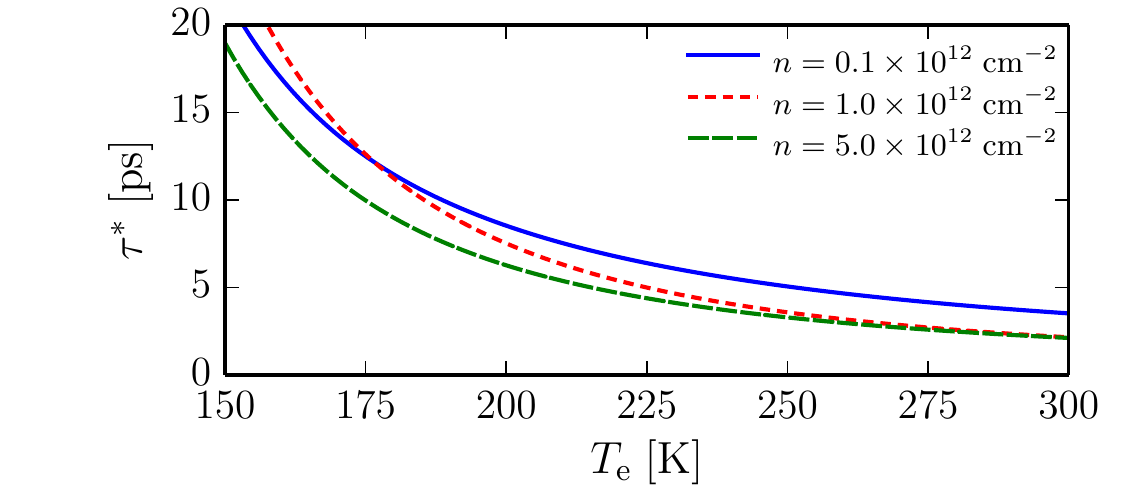}\put(2,42){(b)}
\end{overpic}
\end{tabular}
\end{center}
\caption{(Color online) Panel (a) The cooling time $\tau^\ast$ (\ref{eq:cooling_time}) as a function of the electron density $n$, for different values of the electron temperature $T_{\rm e}$.  Panel (b) The cooling time $\tau^\ast$ as a function of $T_{\rm e}$ for different values of $n$.\label{fig:two}}
\end{figure}

We finally {\it bootstrap} Eq.~(\ref{eq:Q_rate_def}) by introducing  dynamical screening at the level of the random phase approximation (RPA)~\cite{Giuliani_and_Vignale}. This is easily accomplished by the formal replacement
\begin{eqnarray}\label{eq:screening}
{\rm Im}[\chi^{(0)}_{j_\alpha,j_\beta}({\bm q},\omega)] &\to&
\frac{{\rm Im}[\chi^{(0)}_{\rm L}(q,\omega)]}{|\varepsilon(q,\omega)|^2}
\frac{q_{\alpha} q_{\beta}}{q^2} \nonumber\\
&+&{\rm Im}[\chi^{(0)}_{\rm T}(q,\omega)]\left(\delta_{\alpha\beta} - \frac{q_{\alpha} q_{\beta}}{q^2}\right)~,
\end{eqnarray}
where $\chi^{(0)}_{\rm L}(q,\omega)$ and $\chi^{(0)}_{\rm T}(q,\omega)$ are the longitudinal and transverse current-current response functions, and $\varepsilon(q,\omega) = 1 - V_{{\bm q},\omega}\chi^{(0)}_{nn}(q,\omega)$, with $\chi^{(0)}_{nn}(q,\omega) = q^2 \chi^{(0)}_{\rm L}(q,\omega)/\omega^2$ the density-density response function~\cite{density}, is the dynamical RPA screening function~\cite{tomadin_prl_2015}.

Using Eq.~(\ref{eq:screening}) in Eq.~(\ref{eq:Q_rate_def}) and restoring $\hbar$ and $k_{\rm B}$, we finally find the desired expression for the energy transfer rate: 
\begin{equation}\label{eq:Q_rate_fin}
{\cal Q} = 
\frac{\hbar}{4} \int\frac{d^2{\bm q}}{(2\pi)^2}\int_{-\infty}^\infty \frac{d\omega}{\pi} \omega 
[n_{\rm B}(\omega_{\rm e}) - n_{\rm B}(\omega_{\rm L})] {\cal Z}(q,\omega)
~,
\end{equation}
where $\omega_{{\rm e}, {\rm L}}\equiv \hbar \omega/(k_{\rm B} T_{{\rm e}, {\rm L}})$ and
\begin{eqnarray}\label{eq:Z_factor}
{\cal Z}(q,\omega) \equiv 4 \frac{\Im m\left[V_{{\bm q},\omega}\right] {\rm Im}[\chi^{(0)}_{nn}(q,\omega)]}{|\varepsilon(q,\omega)|^2}~.
\end{eqnarray}
A color plot of the real function ${\cal Z}(q,\omega)$ for typical values of microscopic parameters is reported in Fig.~\ref{fig:one}. Eqs.~(\ref{eq:Q_rate_fin})-(\ref{eq:Z_factor}) are the most important results of this work. The transverse part of the current-current response function in Eq.~(\ref{eq:final_propagator}) drops out of the problem since the non-retarded  2D propagator (\ref{eq:final_propagator}) is purely longitudinal.  We note that the quantity ${\cal Z}(q,\omega)$ in Eq.~(\ref{eq:Z_factor}) is dimensionless and bounded, $0\leq {\cal Z}(q,\omega) \leq 1$. The super-Planckian nature of the energy transfer rate (\ref{eq:Q_rate_fin}) stems from contributions to the integral coming from phonon-polariton modes with $q \gg \omega/c$, the only natural short-wavelength cut-off ($\sim k_{\rm F}$) for the integral being provided by the graphene response function ${\rm Im}[\chi^{(0)}_{nn}(q,\omega)]$.

\begin{figure}[t]
\begin{center}
\begin{tabular}{c}
\begin{overpic}[width=1.0\columnwidth]{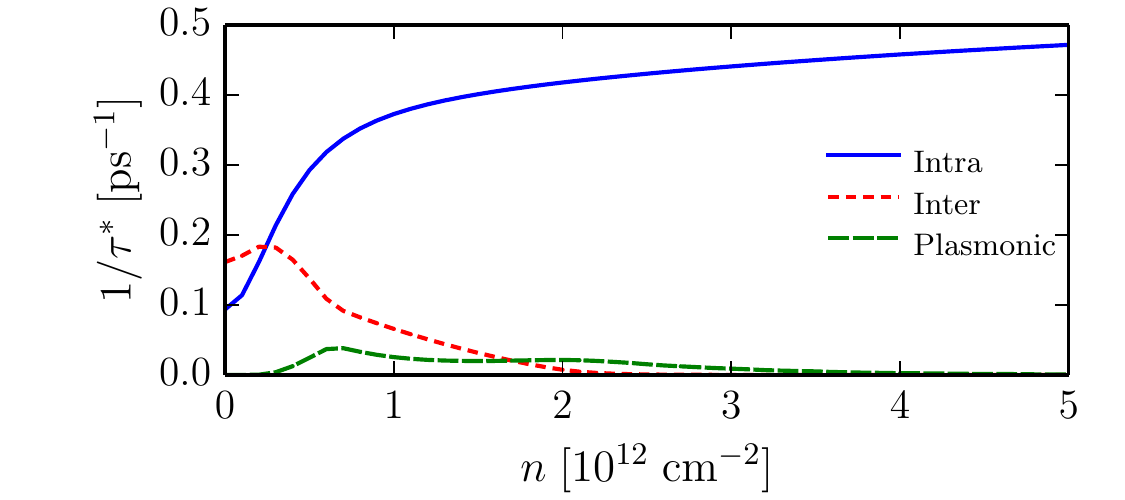}\put(2,42){(a)}
\end{overpic}\\
\begin{overpic}[width=1.0\columnwidth]{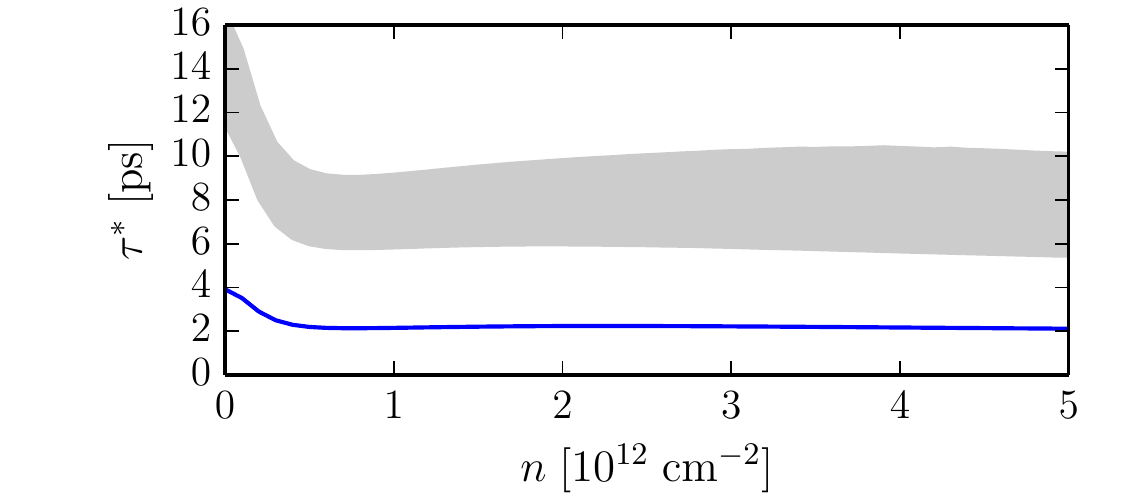}\put(2,42){(b)}
\end{overpic}
\end{tabular}
\end{center}
\caption{(Color online) Panel (a) Intra-band (solid line), inter-band (short-dashed line), and plasmon-phonon polariton (long-dashed line) contributions to the cooling rate $1/\tau^\ast$. The separate contributions to $1/\tau^{\ast}$ are plotted as functions of electron density. Panel (b) A comparison between cooling times for RHT into hBN (blue line) and a non-hyperbolic crystal with identical phonon properties (grey-shaded area). RHT into hyperbolic phonon polaritons is clearly much more efficient. All data for $\tau^\ast$ have been calculated by setting $T_{\rm e} = T_{\rm L} = 300~{\rm K}$.\label{fig:three}}
\end{figure}

{\it Cooling time and temperature dynamics.---}The temperature dynamics $T_{\rm e}(t)$ can be calculated from the differential equation (\ref{eq:EOM}), provided that we introduce the heat capacity. Indeed, using that ${\cal E} =\sum_{{\bm k}, \lambda}(\varepsilon_{{\bm k},\lambda} - \mu)f_{{\bm k}, \lambda}$, we can explicitly calculate $\partial_t{\cal E}$. We find $\partial_t {\cal E}= C_{n} \partial_t T_{\rm e}$, where 
\begin{eqnarray} \label{eq:heat_capacity}
C_{n} &=&
\sum_{{\bm k},\lambda} \left(-\frac{\partial f_{{\bm k},\lambda}}{\partial \varepsilon_{{\bm k},\lambda}}\right) \frac{\big[\varepsilon_{{\bm k},\lambda}-\mu(T_{\rm e}) \big]^2}{k_{\rm B} T_{\rm e}^2} 
\nonumber\\
&+& \frac{\partial \mu(T_{\rm e})}{\partial T_{\rm e}} \sum_{{\bm k},\lambda} \left(-\frac{\partial f_{{\bm k},\lambda}}{\partial \varepsilon_{{\bm k},\lambda}}\right) \frac{\varepsilon_{{\bm k},\lambda}-\mu(T_{\rm e})}{k_{\rm B} T_{\rm e}} 
\end{eqnarray}
is the heat capacity at a constant density $n$. Using Eq.~(\ref{eq:heat_capacity}), we can rewrite Eqs.~(\ref{eq:EOM})-(\ref{eq:Q_rate_fin}) as $C_{n} \partial_t T_{\rm e} = - {\cal Q}$. We now note that we can formally rewrite the latter equation as 
\begin{eqnarray} \label{eq:energy_EOM_3}
\partial_t T_{\rm e}= - \frac{T_{\rm e} - T_{\rm L}}{\tau(T_{\rm e}, T_{\rm L})}~,
\end{eqnarray}
provided that we introduce $\tau(T_{\rm e},T_{\rm L}) \equiv C_{n}(T_{\rm e} - T_{\rm L})/{\cal Q}$. Eq.~(\ref{eq:energy_EOM_3}) can be integrated numerically, as we will discuss below. However, in the limit $\Delta T = T_{\rm e} - T_{\rm L} \to 0$, one can expand the energy transfer rate ${\cal Q}$ for small values of $\Delta T$ and the ratio $(T_{\rm e} - T_{\rm L})/{\cal Q}$ does {\it not} depend on $T_{\rm e}$, i.e.
\begin{eqnarray} \label{eq:cooling_time}
\tau^\ast\equiv \lim_{\Delta T  \to 0}\tau(T_{\rm e}, T_{\rm L}) = \left( \left.\frac{1}{C_{n}} \frac{\partial {\cal Q}}{\partial T_{\rm e}} \right|_{T_{\rm e} = T_{\rm L}} \right)^{-1}~.
\end{eqnarray}
In this case, Eq.~(\ref{eq:energy_EOM_3}) implies a simple exponential decay, $T_{\rm e}(t) = T_{\rm e}(0)\exp(-t/\tau^\ast)$ and $\tau^\ast$ acquires the physical meaning of a {\it cooling time} scale. 

Numerical results for $\tau^\ast$ are shown in Figs.~\ref{fig:two} and~\ref{fig:three}. In particular, in Fig.~\ref{fig:two}(a) we plot  $\tau^\ast$ as a function of carrier density $n$, for different values of the electron temperature $T_{\rm e} = T_{\rm L}$, while in Fig.~\ref{fig:two}(b) we plot  $\tau^\ast$ as a function of $T_{\rm e}$, for different values of $n$.
Note that, in the limit of $n\to 0$, $\tau^\ast$ saturates to a constant, since, in the same limit, the heat capacity converges to a function that depends only on the electron temperature ($C_{n}\propto T^2_{\rm e}$) and so does $\partial Q/\partial T_{\rm e}$. For most values of the electron density away from the $n=0$ charge-neutrality point (CNP), $\tau^\ast$ shows a weak dependence on $n$, because of a cancellation that we now proceed to discuss. Because the integrand in Eq.~(\ref{eq:Q_rate_fin}) is proportional to ${\rm Im}[\chi^{(0)}_{nn}(q,\omega)]$, we can separate out contributions to ${\cal Q}$ that are due to intra-band (i.e.~$\omega<v_{\rm F} q$) and inter-band excitations (i.e.~$\omega>{\rm max}(v_{\rm F} q, 2\varepsilon_{\rm F}/\hbar - v_{\rm F} q)$). There is also a contribution due to plasmon-phonon polaritons (zeroes of $\varepsilon(q,\omega)$, Ref.~\onlinecite{tomadin_prl_2015}), which we define by considering contributions to the frequency integral in Eq.~(\ref{eq:Q_rate_fin}) coming from the $T=0$ Pauli-blocking gap, i.e.~$v_{\rm F} q < \omega < 2\varepsilon_{\rm F}/\hbar - v_{\rm F} q$. These three contributions to the cooling rate $1/\tau^\ast$ are shown in Fig.~\ref{fig:three}(a). We clearly see that the intra-band contribution is dominant for most values of the carrier density, with the exclusion of the low-density regime, where intra- and inter-band contributions become comparable in magnitude. Note also that the increase with $n$ of the intra-band contribution is nearly exactly cancelled by a decrease with $n$ of the inter-band contribution. This explains the weak dependence of $\tau^\ast$ on $n$ away from the CNP displayed in Figs.~\ref{fig:two}(a). The contribution due to the plasmon-phonon polariton branch is negligible. In Fig.~\ref{fig:three}(b), we show the cooling efficiency of the process investigated in this work by comparing RHT into hyperbolic phonon-polaritons (solid line) to RHT into non-hyperpolic phonon-polaritons (grey-shaded area).  The latter is calculated by using Eqs.~(\ref{eq:Q_rate_fin}), (\ref{eq:Z_factor}), and~(\ref{eq:cooling_time}) one time with $\epsilon_x(\omega) \to \epsilon_z(\omega)$ in the equation for $V_{{\bm q},\omega}$, and one time with $\epsilon_z(\omega) \to \epsilon_x(\omega)$. These replacements make sure that the crystal slabs surrounding graphene are non-hyperbolic. We clearly see that RHT into standing hyperbolic phonon-polariton modes is far more efficient.

\begin{figure}[t]
\begin{center}
\includegraphics[width=1.0\columnwidth]{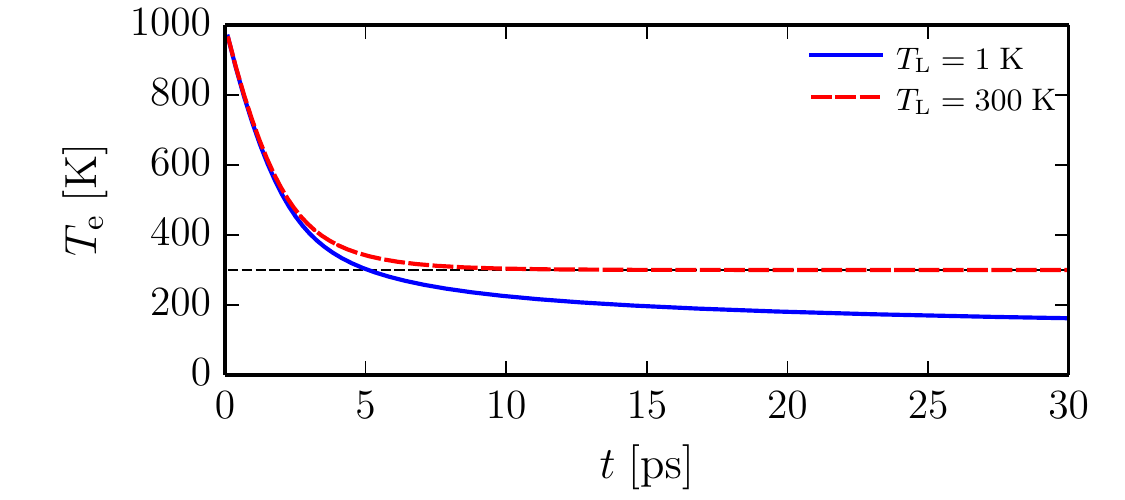}
\end{center}
\caption{(Color online) Cooling dynamics into hyperbolic phonon-polaritons.  We present typical results of the numerical solution of the complete heat equation~(\ref{eq:energy_EOM_3}), where we retained the full dependence of $\tau(T_{\rm e}, T_{\rm L})$ on $T_{\rm e}(t)$. These data have been obtained by setting $T_{\rm e}(0) = 1000~{\rm K}$, $T_{\rm L} = 1~{\rm K}$ (solid line) and $T_{\rm L} = 300~{\rm K}$ (dashed line), and $n=10^{12}~{\rm cm}^{-2}$. We clearly see that, for $T_{\rm L} = 300~{\rm K}$, equilibration with the lattice occurs exponentially fast.\label{fig:four}}
\end{figure}

Before concluding, we would like to discuss temperature dynamics in the overheating $T_{\rm e} \gg T_{\rm L}$ regime. As we have seen above, for $T_{\rm e} \simeq T_{\rm L}$ the function $T_{\rm e}(t)$ is an exponential with time scale $\tau^\ast$. This exponentially fast equilibration does not occur, however, for $T_{\rm e} \gg T_{\rm L}$. In this case, the temperature dynamics $T_{\rm e}(t)$ can be found by solving Eq.~(\ref{eq:energy_EOM_3}) with an initial condition, $T_{\rm e}(0)$. In Fig.~\ref{fig:four}(a) we show that the solution of this equation for $T_{\rm e}(0)=1000~{\rm K} \gg T_{\rm L}=1~{\rm K}$ (solid line) displays a slow decay. Note that, even after $30~{\rm ps}$, the electrons are not equilibrated with the lattice. On the contrary, for $T_{\rm L}=300~{\rm K}$, the dynamics is exponential (dashed line).

In summary, we have presented a theory of near-field thermal radiation transfer between hot carriers in graphene and hyperbolic phonon-polaritons in hBN slabs. Our theory is relevant for understanding the cooling dynamics in ultra-clean encapsulated samples~\cite{geim_nature_2013,mayorov_nanolett_2011,mayorov_nanolett_2012,wang_science_2013,taychatanapat_naturephys_2013,woessner_naturemater_2015,bandurin_science_2016}, where extrinsic mechanisms~\cite{betz_naturephys_2012,graham_naturephys_2013} due to disorder are expected not to be at play. We have discovered that hyperbolic phonon-polaritons in hBN are extremely efficient heat sinks for hot carriers in graphene at room temperature, leading to picosecond time scales for the carrier cooling dynamics in graphene. Within the realm of high-quality samples, this understanding offers a pathway to tuning cooling times by the hBN thickness, which controls the standing phonon-polaritons shown in Fig.~\ref{fig:one}. Thinner hBN slabs tend to lengthen the cooling time, which is a relevant direction for greatly improving the sensitivity of photodetectors~\cite{koppens_naturenano_2014}.

\acknowledgments
This work was supported by the European Union's Horizon 2020 research and innovation programme under grant agreement No.~696656 ``GrapheneCore1'', Fondazione Istituto Italiano di Tecnologia, the Spanish Ministry of Economy and Competitiveness through the ``Severo Ochoa'' Programme for Centres of Excellence in R\&D (SEV-2015-0522), Fundacio Cellex Barcelona, the Mineco grants Ram\'{o}n y Cajal (RYC-2012-12281), Plan Nacional (FIS2013-47161-P), a ``Young Investigator Grant'' (FIS2014-59639-JIN), the Government of Catalonia trough the SGR grant (2014-SGR-1535), the ERC StG ``CarbonLight'' (307806), and the ERC AdG  ``FEMTO/NANO'' (338957). M.B.L. wishes to thank M. Jablan for useful conversations. M.P. is extremely grateful for the financial support granted by ICFO during a visit in August 2016.

\appendix

\section{Electrostatics of the hBN/Graphene/hBN vertical stack}
\label{sect:Appendix}
\renewcommand\thefigure{A\arabic{figure}}
\renewcommand\theequation{A\arabic{equation}}
\setcounter{equation}{0}

We consider a vertical heterostructure composed of: 1) a graphene sheet located at $z = 0$; 
2) a homogeneous but uniaxial insulator of thickness $d^\prime$ with dielectric tensor 
$\hat{\bm \epsilon} = {\rm diag}(\epsilon_{x},\epsilon_{x},\epsilon_{z})$ placed {\it above} graphene; 
3)  a homogeneous but uniaxial insulator of thickness $d$ with dielectric tensor 
$\hat{\bm \epsilon} = {\rm diag}(\epsilon_{x},\epsilon_{x},\epsilon_{z})$ placed {\it below} graphene; 
4) homogeneous and isotropic insulators with dielectric constants $\epsilon_{\rm a}$ and $\epsilon_{\rm b}$ filling the two half-spaces $z>d^\prime$ and $z < -d$, respectively.

We calculate the electrical potential created by an electron sitting at the origin in graphene by following the approach of Ref.~\onlinecite{tomadin_prl_2015}. The three-dimensional displacement field ${\bm D}({\bm r}, z)$ in cylindrical coordinates must satisfy the condition ${\bm \nabla} \cdot {\bm D}({\bm r}, z) =0$ everywhere in space. However, the presence of an electron with charge density $-e \delta^2({\bm r})\delta(z)$ at $z=0$ implies a discontinuity of the normal component $D_{z}$ of the displacement field across $z=0$, while the tangential components $E_{x}, E_{y}$ of the electric field ${\bm E}({\bm r}, z)$ must be continuous.

Since the electric field ${\bm E}({\bm r}, z)$ is irrotational everywhere in space, we can introduce the electric potential $\phi({\bm r}, z)$ in the four regions of space $z>d^\prime$, $d^\prime<z<0$, $-d < z < 0$, and $z < -d$. The Laplace equation $- \epsilon_{x}\partial^2_x \phi({\bm r}, z) - \epsilon_{y}\partial^2_y \phi({\bm r}, z)- \epsilon_{z}\partial^2_z \phi({\bm r}, z) =0$ in the uniaxial dielectrics (i.e.~for $d^\prime<z<0$ and $-d < z<0$) can be reduced~\cite{Landau08} to an ordinary Laplace equation by scaling $x \to x/\sqrt{\epsilon_{x}}$, $y \to y/\sqrt{\epsilon_{y}}$, and $z \to z/\sqrt{\epsilon_{z}}$. 

\begin{table}[t]
\begin{ruledtabular}
\begin{tabular}{l | c c}
\, & $\ell=x$  & $\ell=z$	\\
\hline
$\epsilon_{\ell,0}$	& 6.70 & 3.56 \\
$\epsilon_{\ell,\infty}$ & 4.87 & 2.95 \\	
$\gamma_\ell~[{\rm meV}]$ & 0.87 & 0.25 \\
$\hbar\omega_{\ell}^{\rm T}~[{\rm meV}]$ & 170.1 & 92.5 \\
$\hbar\omega_{\ell}^{\rm L}~[{\rm meV}]$ & 199.5 & 101.6 \\
\end{tabular}
\end{ruledtabular}
\caption{Microscopic parameters entering Eq.~(\ref{eq:BN_dielectric}). See also Supplementary Information in Ref.~\onlinecite{woessner_naturemater_2015}.\label{table_parameters}}
\end{table}

Imposing the aforementioned boundary conditions and carrying out tedious but elementary algebraic steps, we find the following expression for 
the dressed 2D Coulomb interaction on the real-frequency axis:
\begin{widetext}
\begin{equation}\label{eq:effective_interaction_uniaxial}
\begin{split}
&V_{{\bm q},\omega}=v_q \frac{1}{2} \Bigg\{ \sqrt{\epsilon_x\epsilon_z} + (\epsilon_{\rm a} + \epsilon_{\rm b})
\tanh{\left[ q\sqrt{\frac{\epsilon_x}{\epsilon_z}}(d+d^\prime) \right]} + (\epsilon_{\rm b} - \epsilon_{\rm a})
\frac{\sinh{\left[ q\sqrt{\frac{\epsilon_x}{\epsilon_z}}(d - d^\prime) \right]}}{\cosh{\left[ q\sqrt{\frac{\epsilon_x}{\epsilon_z}}(d + d^\prime) \right]}} \\
&+\left( \sqrt{\epsilon_x\epsilon_z} - \frac{\epsilon_{\rm a} \epsilon_{\rm b}}{\sqrt{\epsilon_x\epsilon_z}} \right) 
\frac{\cosh{\left[ q\sqrt{\frac{\epsilon_x}{\epsilon_z}}(d^\prime - d) \right]}}{\cosh{\left[ q\sqrt{\frac{\epsilon_x}{\epsilon_z}}(d + d^\prime) \right]}} + \frac{\epsilon_{\rm a}\epsilon_{\rm b}}{\sqrt{\epsilon_x\epsilon_z}} \Bigg\}
 \Bigg\{\sqrt{\epsilon_x\epsilon_z} + {\tilde \epsilon}
\tanh{\left[ q\sqrt{\frac{\epsilon_x}{\epsilon_z}}(d + d^\prime) \right]}  \Bigg\}^{-1}
\end{split}
\end{equation}
\end{widetext}
where $v_q = 4\pi e^2 /[q(\epsilon_{\rm a} +\epsilon_{\rm b})]$ and 
${\widetilde \epsilon} \equiv (\epsilon_{\rm a}\epsilon_{\rm b} + \epsilon_x\epsilon_z)/(\epsilon_{\rm a} + \epsilon_{\rm b})$. In the limit $d^\prime \to 0$, Eq.~(\ref{eq:effective_interaction_uniaxial}) reduces to a recently derived result~\cite{tomadin_prl_2015}.

The frequency dependence of the dressed Coulomb interaction $V_{{\bm q},\omega}$ is due to optical phonons in the hBN slabs~\cite{tomadin_prl_2015}. Indeed, in the case of hBN, the components of the uniaxial dielectric tensor have an important dependence on frequency in the mid infrared, which is usually parametrized in the following form
\begin{eqnarray} \label{eq:BN_dielectric}
\epsilon_\ell(\omega) &=& \epsilon_{\ell,\infty} + \frac{\epsilon_{\ell,0} - \epsilon_{\ell,\infty}}{1 - (\omega/\omega_{\ell}^{\rm T})^2 + i \gamma_\ell \hbar \omega/(\hbar \omega_{\ell}^{\rm T})^2}~,
\end{eqnarray}
with $\ell = x$ or $z$.  Here $\epsilon_{\ell,0}$ and $\epsilon_{\ell, \infty}$ are the static and high-frequency dielectric constants, respectively, while $\omega^{\rm T}_{\ell}$ is the transverse optical phonon frequency in the direction $\ell$. The longitudinal optical phonon frequency $\omega^{\rm L}_{\ell}$ satisfies the Lyddane-Sachs-Teller relation $\omega^{\rm L}_{\ell} = \omega_{\ell}^{\rm T} \sqrt{\epsilon_{\ell,0}/\epsilon_{\ell,\infty}}$. The parameter $\gamma_{\ell}$ represents hBN phonon losses (in~${\rm meV}$).

All the calculations reported in the main text have been done with the dielectric parameters reported in Table~\ref{table_parameters}. Also, we have taken $\epsilon_{\rm a} =1$, $\epsilon_{\rm b} = 3.9$, $d^\prime=9~{\rm nm}$, and $d=27~{\rm nm}$.


\begin{thebibliography}{77}
%
\bibitem{hotcarriers}
{\it Hot Carriers in Semiconductors}, edited by J. Shah and G.J. Iafrate  (Academic, London, 1992).
%
\bibitem{bistritzer_prl_2009}
R. Bistritzer and A.H. MacDonald, \href{http://dx.doi.org/10.1103/PhysRevLett.102.206410}{Phys. Rev. Lett.~{\bf 102}, 206410 (2009)}.
%
\bibitem{tse_prb_2009}
W.-K. Tse and S. Das Sarma, \href{http://dx.doi.org/10.1103/PhysRevB.79.235406}{Phys. Rev. B~{\bf 79}, 235406 (2009)}.
%
\bibitem{koppens_naturenano_2014}
K.H.L. Koppens, T. Mueller, Ph. Avouris, A.C. Ferrari, M.S. Vitiello, and M. Polini, 
\href{http://dx.doi.org/10.1038/nnano.2014.215}{Nature Nanotech.~{\bf 9}, 780 (2014)}.
%
\bibitem{geim_naturemater_2007} 
A.K. Geim and K.S. Novoselov, \href{http://dx.doi.org/10.1038/nmat1849}{Nature Mater.~{\bf 6}, 183 (2007)}.
%
\bibitem{betz_naturephys_2012}
A.C. Betz, S.H. Jhang, E. Pallecchi, R. Ferreira, G. F\`{e}ve, J-M. Berroir, and B. Pla\c{c}ais, 
\href{http://dx.doi.org/10.1038/nphys2494}{Nature Phys.~{\bf 9}, 109 (2012)}.
%
\bibitem{graham_naturephys_2013}
M.W. Graham, S.-F. Shi, D.C. Ralph, J. Park, and P.L. McEuen, 
\href{http://dx.doi.org/10.1038/nphys2493}{Nature Phys.~{\bf 9}, 103 (2013)}.
%
\bibitem{song_prl_2012}
J.C.W. Song, M.Y. Reizer, and L.S. Levitov, 
\href{http://dx.doi.org/10.1103/PhysRevLett.109.106602}{Phys. Rev. Lett.~{\bf 109}, 106602 (2012)}.
%
\bibitem{song_jpcm_2015}
J.C.W. Song and L.S. Levitov, \href{http://dx.doi.org/10.1088/0953-8984/27/16/164201}{J. Phys.: Condens. Matter~{\bf 27}, 164201 (2015)}.
%
\bibitem{geim_nature_2013}
A.K. Geim and I.V. Grigorieva, \href{http://dx.doi.org/10.1038/nature12385}{Nature~{\bf 499}, 419 (2013)}.
%
\bibitem{mayorov_nanolett_2011}
A.S. Mayorov, R.V. Gorbachev, S.V. Morozov, L. Britnell, R. Jalil, L.A. Ponomarenko, P. Blake, K.S. Novoselov, K. Watanabe, T. Taniguchi, and A.K. Geim, \href{http://dx.doi.org/10.1021/nl200758b}{Nano Lett.~{\bf 11}, 2396 (2011)}.
%
\bibitem{mayorov_nanolett_2012}
A.S. Mayorov, D.C. Elias, I.S. Mukhin, S.V. Morozov, L.A. Ponomarenko, K.S. Novoselov, A.K. Geim, and R.V. Gorbachev, \href{http://dx.doi.org/10.1021/nl301922d}{Nano Lett.~{\bf 12}, 4629 (2012)}.
%
\bibitem{wang_science_2013}
L. Wang, I. Meric, P.Y. Huang, Q. Gao, Y. Gao, H. Tran, T. Taniguchi, K. Watanabe, L.M. Campos, D.A. Muller, J. Guo, P. Kim, J. Hone, K.L. Shepard, and C.R. Dean, \href{http://dx.doi.org/10.1126/science.1244358}{Science~{\bf 342}, 614 (2013)}.
%
\bibitem{taychatanapat_naturephys_2013}
T. Taychatanapat, K. Watanabe, T. Taniguchi, and P. Jarillo-Herrero, \href{http://dx.doi.org/10.1038/nphys2549}{Nature Phys.~{\bf 9}, 225 (2013)}.
%
\bibitem{woessner_naturemater_2015}
A. Woessner, M.B. Lundeberg, Y. Gao, A. Principi, P. Alonso-Gonz\'alez, M. Carrega, K. Watanabe, T. Taniguchi, G. Vignale, M. Polini, J. Hone, R. Hillenbrand, and F.H.L. Koppens, \href{http://dx.doi.org/10.1038/nmat4169}{Nature Mater.~{\bf 14}, 421 (2015)}.
%
\bibitem{bandurin_science_2016}
D. Bandurin, I. Torre, R.K. Kumar, M. Ben Shalom, A. Tomadin, A. Principi, G.H. Auton, E. Khestanova, K.S. NovoseIov, I.V. Grigorieva, L.A. Ponomarenko, A.K. Geim, and M. Polini, \href{http://dx.doi.org/10.1126/science.aad0201}{Science~{\bf 351}, 1055 (2016)}.
%
\bibitem{hwang_prb_2008}
E.H. Hwang and S. Das Sarma, \href{http://dx.doi.org/10.1103/PhysRevB.77.115449}{Phys. Rev. B~{\bf 77}, 115449 (2008)}.
%
\bibitem{principi_prb_2014}
A. Principi, M. Carrega, M.B. Lundeberg, A. Woessner, F.H.L. Koppens, G. Vignale, and M. Polini, \href{http://dx.doi.org/10.1103/PhysRevB.90.165408}{Phys. Rev. B~{\bf 90}, 165408 (2014)}.
%
\bibitem{hyperbolicmaterials}
A. Poddubny, I. Iorsh, P. Belov, and Y. Kivshar, 
\href{http://dx.doi.org/10.1038/nphoton.2013.243}{Nature Photon.~{\bf 7}, 948 (2013)}.
%
\bibitem{dai_science_2014}
S. Dai, Z. Fei, Q. Ma, A.S. Rodin, M. Wagner, A.S. McLeod, M.K. Liu, W. Gannett, W. Regan, 
K. Watanabe, T. Taniguchi, M. Thiemens, G. Dominguez, A.H. Castro Neto, A. Zettl, F. Keilmann, 
P. Jarillo-Herrero, M.M. Fogler, and D.N. Basov, \href{http://dx.doi.org/10.1126/science.1246833}{Science~{\bf 343}, 1125 (2014)}.
%
\bibitem{caldwell_naturecommun_2014}
J.D. Caldwell, A. Kretinin, Y. Chen, V. Giannini, M.M. Fogler, Y. Francescato, 
C.T. Ellis, J.G. Tischler, C.R. Woods, A.J. Giles, M. Hong, K. Watanabe, 
T. Taniguchi, S.A. Maier, and K.S. Novoselov, 
\href{http://dx.doi.org/10.1038/ncomms6221}{Nature Commun.~{\bf 5}, 5221 (2014)}.
%
\bibitem{tomadin_prl_2015}
A. Tomadin, A. Principi, J.C.W. Song, L.S. Levitov, and M. Polini, 
\href{http://dx.doi.org/10.1103/PhysRevLett.115.087401}{Phys. Rev. Lett.~{\bf 115}, 087401 (2015)}.
%
\bibitem{hargreaves_pla_1969}
C.M. Hargreaves, \href{http://dx.doi.org/10.1016/0375-9601(69)90264-3}{Phys. Lett. A~{\bf 30},  491 (1969)}.
%
\bibitem{polder_prb_1971}
D. Polder and M. Van Hove, \href{http://dx.doi.org/10.1103/PhysRevB.4.3303}{Phys. Rev. B~{\bf 4}, 3303 (1971)}.
%
\bibitem{rytov_book_1953}
S.M. Rytov, {\it Theory of Electric Fluctuations and Thermal Radiation} (Academy of Sciences of USSR, Moscow, 1953).
%
\bibitem{lifshitz_zetf_1956}
E.M. Lifshitz, \href{http://www.jetp.ac.ru/cgi-bin/dn/e_002_01_0073.pdf}{Sov. Phys. JETP~{\bf 2}, 73 (1956)}.
%
\bibitem{dzyaloshinskii_ap_1961}
I.E. Dzyaloshinskii, E.M. Lifshitz, and L.P. Pitaevskii, 
\href{http://dx.doi.org/10.1080/00018736100101281}{Adv. Phys.~{\bf 10}, 165 (1961)}.
%
\bibitem{wilde_nature_2006}
Y. De Wilde, F. Formanek, R. Carminati, B. Gralak, P.-A. Lemoine, K. Joulain, J.-P. Mulet, 
Y. Chen, and J.-J. Greffet, \href{http://dx.doi.org/10.1038/nature05265}{Nature~{\bf 444}, 740 (2006)}.
%
\bibitem{shen_nanolett_2009}
S. Shen, A. Narayanaswamy, and G. Chen, 
\href{http://dx.doi.org/10.1021/nl901208v}{Nano Lett.~{\bf 9}, 2909 (2009)}.
%
\bibitem{rousseau_naturephoton_2011}
E. Rousseau, A. Siria, G. Jourdan, S. Volz, F. Comin, J. Chevrier and J.-J. Greffet, 
\href{http://dx.doi.org/10.1038/nphoton.2009.144}{Nature Photon.~{\bf 3}, 514 (2009)}.
%
\bibitem{ottens_prl_2011}
R.S. Ottens, V. Quetschke, S. Wise, A.A. Alemi, R. Lundock, G. Mueller, D.H. Reitze, D.B. Tanner, 
and B.F. Whiting, \href{http://dx.doi.org/10.1103/PhysRevLett.107.014301}{Phys. Rev. Lett.~{\bf 107}, 014301 (2011)}.
%
\bibitem{jones_nanolett_2012}
A.C. Jones and M.B. Raschke, \href{http://dx.doi.org/10.1021/nl204201g}{Nano Lett.~{\bf 12}, 1475 (2012)}.
%
\bibitem{kim_nature_2015}
K. Kim, B. Song, V. Fern\'{a}ndez-Hurtado, W. Lee, W. Jeong, L. Cui, D. Thompson, J. Feist, 
M.T. Homer Reid, F.J. Garc\'{i}a-Vidal, J.C. Cuevas, E. Meyhofer, and P. Reddy, 
\href{http://dx.doi.org/10.1038/nature16070}{Nature~{\bf 528}, 387 (2015)}.
%
\bibitem{pendry_jpcm_1999}
J.B. Pendry, \href{http://dx.doi.org/10.1088/0953-8984/11/35/301}{J. Phys.: Condens. Matt.~{\bf 11}, 6621 (1999)}.
%
\bibitem{volokitin_prb_2001}
A.I. Volokitin and B.N.J. Persson, 
\href{http://dx.doi.org/10.1103/PhysRevB.63.205404}{Phys. Rev. B~{\bf 63}, 205404 (2001)}.
%
\bibitem{benabdallah_prb_2010}
P. Ben-Abdallah and K. Joulain, \href{http://dx.doi.org/10.1103/PhysRevB.82.121419}{Phys. Rev. B~{\bf 82}, 121419 (2010)}.
%
\bibitem{guo_apl_2012}
Y. Guo, C.L. Cortes, S. Molesky, and Z. Jacob, \href{http://dx.doi.org/10.1063/1.4754616}{Appl. Phys. Lett.~{\bf 101}, 131106 (2012)}.
%
\bibitem{messina_prb_2013}
R. Messina, M. Tschikin, S.-A. Biehs, and P. Ben-Abdallah, 
\href{http://dx.doi.org/10.1103/PhysRevB.88.104307}{Phys. Rev. B~{\bf 88}, 104307 (2013)}.
%
\bibitem{benabdallah_prl_2014}
P. Ben-Abdallah and S.-A. Biehs, \href{http://dx.doi.org/10.1103/PhysRevLett.112.044301}{Phys. Rev. Lett.~{\bf 112}, 044301 (2014)}.
%
\bibitem{liu_acsphoton_2014}
X.L. Liu, R.Z. Zhang, and Z.M. Zhang, 
\href{http://dx.doi.org/10.1021/ph5001633}{ACS Photonics~{\bf 1}, 785 (2014)}.
%
\bibitem{miller_prl_2015}
O.D. Miller, S.G. Johnson, and A.W. Rodriguez, 
\href{http://dx.doi.org/10.1103/PhysRevLett.115.204302}{Phys. Rev. Lett.~{\bf 115}, 204302 (2015)}.
%
\bibitem{kotov_rmp_2012}
V.N. Kotov, B. Uchoa, V.M. Pereira, F. Guinea, and A.H. Castro Neto, \href{http://dx.doi.org/10.1103/RevModPhys.84.1067}{\rmp~{\bf 84}, 1067 (2012)}.
%
\bibitem{low_prb_2012}
T. Low, V. Perebeinos, R. Kim, M. Freitag, and Ph. Avouris, 
\href{http://dx.doi.org/10.1103/PhysRevB.86.045413}{Phys. Rev. B~{\bf 86}, 045413 (2012)}.
%
\bibitem{principi_prb_2009}
A. Principi, M. Polini, and G. Vignale, 
\href{http://dx.doi.org/10.1103/PhysRevB.80.075418}{Phys. Rev. B~{\bf 80}, 075418 (2009)}.
%
\bibitem{Giuliani_and_Vignale}
G.F. Giuliani and G. Vignale, {\it Quantum Theory of the Electron Liquid} 
(Cambridge University Press, Cambridge, 2005).
%
\bibitem{density}
B. Wunsch,  T. Stauber, F. Sols, and F. Guinea, \href{http://dx.doi.org/10.1088/1367-2630/8/12/318}{New J. Phys. {\bf 8}, 318 (2006)}; E.H. Hwang and S. Das Sarma, \href{http://dx.doi.org/10.1103/PhysRevB.75.205418}{\prb~{\bf 75}, 205418 (2007)}.
%
\bibitem{Landau08}
L.D. Landau and E.M. Lifshitz, {\it Course of Theoretical Physics: Electrodynamics of Continuos Media} (Pergamon, New York, 1984).
%
\end{thebibliography}
\end{document}